# Epitaxial thin films of multiferroic $Bi_2FeCrO_6$ with B-site cationic order

Riad Nechache[1], Louis-Philippe Carignan[3], Lina Gunawan[2], Catalin Harnagea[1], Gianluigi A. Botton[2], David Ménard[3], and Alain Pignolet[1]

[1] Prof. A. Pignolet, C. Harnagea, R. Nechache,
   INRS Énergie, Materiaux et Télécommunications,
   1650, boulevard Lionel-Boulet,
   Varennes (Québec), J3X 1S2 (Canada).
   E-mail : pignolet@emt.inrs.ca

[2] Prof. G. A. Botton, L. Gunawan
    Dept of Materials Science and Engineering
    Brockhouse Institute for Materials Research and Centre for Emerging Device Technologies
    McMaster University
    1280 Main Street West
    Hamilton( Ontario), L8S 4M1 (Canada).
    E-Mail: gbotton@mcmaster.ca

[3] Prof D. Ménard, L.-P. Carignan
    École Polytechnique de Montréal,
    Département de Génie Physique,
    P.O. Box 6079, Station. Centre-ville,
    Montréal, (Québec), H3C 6A7 (Canada).
    E-mail: david.menard@polymtl.ca

**Abstract**
Epitaxial thin films of $Bi_2FeCrO_6$ have been synthesized by pulsed laser deposition on $SrRuO_3$ on (100)- and (111)-oriented $SrTiO_3$ substrates. Detailed X-ray diffraction and cross-section transmission electron microscopy analysis revealed a double perovskite crystal structure of the $Bi_2FeCrO_6$ epitaxial films very similar to that of $BiFeO_3$ along with a particularly noteworthy $Fe^{3+}/Cr^{3+}$ cation ordering along the [111] direction. The films contain no detectable magnetic iron oxide impurities and have the correct cationic average stoichiometry throughout their thickness. They however exhibit a slight modulation in the Fe and Cr compositions forming complementary stripe patterns, suggesting minor local excess or depletion of Fe and Cr.
The epitaxial BFCO films exhibit good ferroelectric and piezoelectric properties, in addition to magnetic properties *at room temperature*, as well as an unexpected crystallographic orientation dependence of their room temperature magnetic properties.
Our results qualitatively confirm the predictions made using the *ab-initio* calculations: the double-perovskite structure of $Bi_2FeCrO_6$ films exhibit a $Fe^{3+}/Cr^{3+}$ cation ordering and good multiferroic properties, along with the unpredicted existence of magnetic ordering at room temperature.

**I. Introduction**

Multiferroics, defined as materials which combine at least two "ferroic" properties such as ferromagnetic order (spontaneous magnetic polarization that can be reversed by a magnetic field), ferroelectricity (e.g. spontaneous electric polarization that can be switched by an applied electric field) or ferroelasticity (change in electric polarization accompanied by a change in shape), have recently seen a surge of research activity.[1,2] A subset among the multiferroics are the *ferroelectromagnets*, which possess both electric and magnetic order parameters. Beside their exciting physics, multiferroics could bring innovative solutions for applications in many fields,[3] such as agile electromagnetics, optoelectronics or spintronics.[4] The *magnetoelectric coupling* that exists in *ferroelectromagnets*,[2,5] could allow for magnetization reversal by applying an electric field[6,7] instead of a magnetic field. It may arise directly between the magnetic and electric order parameters or indirectly via elastic strain. In her seminal paper discussing the conditions required for ferroelectricity and ferromagnetism to coexist in oxides, N. A. Hill stressed that such conditions were rarely met in nature.[1]

One of the known natural ferroelectromagnet is $BiFeO_3$ (BFO), which has ferroelectric (with $T_C$ ~1100K) and antiferromagnetic ($T_N$ ~640K) properties at room temperature. Even though the bulk samples have been synthesized back in 1950s, characterizations of its intrinsic properties have been difficult due to poor sample quality. A few years ago, a gigantic enhancement of the magnetic moment in thin films of BFO (compared to the magnetic moment in bulk) was reported by Wang *et al*.[8] However, some controversy over these results appeared shortly after publication and the intrinsic magnetic properties of BFO films are still debated.[9,10] Recently Béa et *al.* found that phase pure BFO films showing no indications of parasitic Fe-rich phases have a low bulk-like magnetic moment (~0.02 $\mu_B$/Fe) while films containing parasitic Fe oxides (mostly the γ-$Fe_2O_3$ maghemite phase) display a weak ferromagnetic behavior.[11] Such an epitaxial mixture of ferroelectric BFO and magnetic Fe oxide represents, however, a new type of multiferroic nanocomposite material.[12] In fact, it has been suggested that both magnetization and polarization could independently encode information in a single multiferroic bit. Such a four-state memory has indeed recently been demonstrated,[13] but in practice it is likely that the two order parameters are coupled.[14,15] This coupling could in principle permit data to be written electrically and read magnetically, or vice-versa, enabling to exploit the best aspect of ferroelectric random access memory (FeRAM) and magnetic data storage (MRAM), while avoiding the problems associated with reading-induced fatigue in FeRAM or large local magnetic field generation needed for magnetic writing (i.e. miniaturized coils difficult to produce and consuming too much power). Although significant material developments are still required to generate magnetoelectric materials that could make a real contribution to the data strorage industry, the study of novel multiferroic materials remains important in order for disruptive technologies ultimately to emerge. In this context, Baettig and Spaldin, using *ab-initio* first-principles density functional theory, reported the design and properties of a new multiferroic material, namely $Bi_2FeCrO_6$ (BFCO).[16] They predicted that BFCO would have a rhombohedral structure very similar to that of BFO, except that every second iron cation in the (111) direction is replaced by a chromium cation. They also predicted that BFCO would be ferrimagnetic at 0 K with a magnetic moment of 2 $\mu_B$ per formula unit and would be ferroelectric with a polarization of 80 $\mu C\ cm^{-2}$.

We recently reported the first synthesis of epitaxial BFCO films by pulsed laser deposition (PLD), and their good multiferroic properties *at room temperature*.[17] The PLD-grown epitaxial BFCO films were deposited directly on (100)-oriented $SrTiO_3$ (STO) single crystalline substrates[18] and on $SrTiO_3$ covered with an epitaxial $SrRuO_3$ (SRO) buffer layer.[17] Our results confirmed the existence of multiferroic properties in BFCO films that were predicted by *ab-initio* calculations. The existence of a magnetic ordering *at room temperature* had not been theoretically predicted then, but is a very promising result that needs to be further investigated.

We report here on the successful growth by PLD of (111)-oriented multiferroic BFCO thin films and present a comparative analysis of the structural, ferroelectric and magnetic properties for (111)-oriented and (100)-oriented BFCO epitaxial films. The films were grown on STO(111) and STO(100) substrates with an intermediate epitaxial buffer layer of the conducting perovskite SrRuO$_3$ (SRO) serving as the bottom electrode for electrical characterization. Both compositional and structural properties of the films as well as their ferroelectric and magnetic properties are presented and discussed. The major outcome of this investigation is the demonstration of the Fe$^{+3}$/Cr$^{+3}$ cationic ordering along the [111] crystallographic direction within the double perovskite unit cell - an important question which remained open in our first reports:[17,18] This precludes the rationalization of our results by the presence of two distinct epitaxial BiFeO$_3$ and BiCrO$_3$ phases.

## II. Experimental

BFCO films were grown on top of 30 nm thick epitaxial layers of SrRuO$_3$ (SRO) on (100)- and (111)-oriented SrTiO$_3$ (STO) single crystalline substrates. Both BFCO and SRO films were synthesized by pulsed laser ablation deposition (PLD) with a KrF excimer laser (248 nm, 8Hz, target-substrate = 5.5 cm). BFCO films were grown from a Bi$_2$FeCrO$_6$ stoichiometric target in an ambient of 9 mTorr O$_2$ at a substrate temperature of 680ºC and a laser fluence of 2.0 J cm$^{-2}$. SRO films were deposited from a SrRuO$_3$ stoichiometric target in 100 mTorr O$_2$ with a fluence of 2.0 J cm$^{-2}$ on substrates heated at 650ºC. The growth rates of BFCO and SRO were 10 nm/min and 6 nm/min, respectively. After deposition, the films were cooled down to 400ºC at 8ºC/min in an oxygen atmosphere and tempered for 1 hour at this temperature, before they were further cooled down to room temperature. The film thicknesses were measured using a variable angle spectroscopic ellipsometer (J. A. Woollam Co) with photon energy between 0.1 and 3 eV. The films having a thickness less than 150 nm were also measured by X-ray reflectometry (XRR) and compared to the ellipsometry measurements (150 nm is the maximum thickness measured by our XRR setup).

The BFCO film stoichiometry was analyzed by Rutherford Backscattering Spectrometry (RBS) (350 keV Van der Graff accelerator) as well as using energy dispersive X-ray spectroscopy (EDXS) in a transmission electron microscope (TEM). The oxidation states of the cations, as well as the composition, were also investigated across the whole layer's depth for a 300 nm thick film using X-ray photoelectron spectroscopy (XPS) (ESCALAB 220i-XL system). Depth profiling was performed with intermittent Ar$^+$ ion sputtering. The sputtering was done with a 3 keV Ar$^+$ ion beam (beam current: 0.1 - 0.2 µA) in 2-3 X10$^{-8}$ Torr of Ar. The ion beam was scanned over an area of 2 × 2 mm$^2$. Under the conditions used, the average sputtering rate measured was about 0.033 nm/sec. The x-ray photoelectron Bi4f, Fe2p, Cr2p and O1s core level spectra of the BFCO successive surface layers were taken with hemispherical electron energy analyzer using a Mg K$_\alpha$ X-ray twin source (1253.6 eV). During XPS measurement, the pressure inside the analysis chamber was maintained below 3 X10$^{-9}$ Torr. The atomic concentration was determined by using the sensitivity factors supplied by the instrument manufacturer. The binding energy scale was corrected and calibrated using the Bi4$f_{7/2}$ peak position (158.8 eV). For comparison, the same measurements were performed on a 300 nm thick BFO film obtained under similar conditions.[19]

Structural characterization of BFCO films was performed using X-ray diffraction (XRD) (PANalytical X'Pert MRD 4-circle diffractometer). High-resolution analytical electron microscopy was carried out with a JEOL 2010F scanning transmission electron microscope equipped with a field emission source, an electron energy loss spectrometer (Gatan GIF), a high-angle annular field (HAADF) detector (Fischione Instruments HAADF detector) and an energy dispersive X-ray spectroscopy (EDXS) detector (Oxford Instrument Pentafet with Inca spectrum analyzer software). Conventional transmission electron microscopy and selected area electron diffraction was carried out with a Philips CM12 TEM.  Elemental maps were obtained with the EDXS method with a probe size of less than 1nm.

Piezoresponse force microscopy (PFM) [17-23] was used to image and manipulate the local ferroelectric properties of the BFCO films. Here we used a DI-Enviroscope AFM (Veeco) equipped with a NSC36a (Micromasch) cantilever and tips coated with Co/Cr. We applied an ac voltage of 0.5V at 26 kHz between the conductive tip and the SRO bottom electrode located beneath the BFCO layer and we detected the induced piezoelectric surface vibrations using a Lock-in Amplifier from Signal Recovery (model 7265).

The magnetization as a function of the magnetic field (M-H hysteresis loops) was measured at room temperature using a vibrating-sample magnetometer (VSM) (Model EV9 from ADE Technologies). The induced magnetization, parallel to the applied field, is measured for fields ranging from -10 KOe to +10 kOe. A maximum saturating field of 10 kOe was applied parallel to film plane, then decreased down to -10 kOe by step of 500 Oe, and back up to 10 kOe.  The field steps were decreased down to 50 Oe around the zero field range for better resolution.  An average factor of 20 measurements per field points provided an absolute sensitivity of about $10^{-6}$ emu. The presence of an ordered magnetic phase is indicated by the observation of a magnetic hysteresis. The magnetic responses of the sample holding rods and of bare substrates (without BFCO) were also measured  and subtracted in order to ensure that the ferro- or ferrimagnetic signal was originating solely from the BFCO films. [24]

### III. Results and Discussion

RBS analysis of the BFCO/SRO/STO(100) heterostructure (spectrum not shown) indicates that the cationic ratio Bi/(Fe+Cr) was close to unity (within experimental resolution of 3%). The Fe/Cr ratio and the oxygen content are difficult to determine due to the close position of the Fe and Cr peaks and the small oxygen scattering cross-section, respectively. However, this information can be found using other chemical analysis techniques, namely X-ray microanalysis in a transmission electron microscope (TEM) and an X-ray photoelectron spectroscopy (XPS) analysis. Both techniques confirmed that the average Fe/Cr ratio is about unity. XPS depth profiling results (Fig. 1A) reveal that the oxidation state of Fe and Cr ions in the film remains 3+ throughout the films thickness and hence exclude (within the detection limit of the technique) the presence of magnetite ($Fe_3O_4$) iron-oxide impurities in our BFCO films. Also absent are the satellite peaks characteristic of $\gamma$-$Fe_2O_3$, thus ruling out maghemite ($\gamma$-$Fe_2O_3$) inclusions. These results are similar to those obtained for BFCO thin films deposited directly on (100)-oriented STO substrates.[18] It follows that the chemical environment observed around the iron cations in films deposited on SRO epitaxial films, on a STO substrate, is the same as that observed in films directly deposited on a STO substrate. This confirms the earlier results that also excluded the presence of magnetic iron oxide phases.[18] An EDXS line profile obtained from cross-sectional TEM image of BFCO/SRO/STO100 heterostructure is presented in Figure 1B. The concentration of Bi,

Fe and Cr cations were found nearly constant across the film thickness and are approximately 20%, 10%, and 10 %, respectively (within experimental error of 3 at%), as expected. Moreover, this result confirms that Fe and Cr ions are homogeneously distributed throughout the depth. We can notice that the SRO layer shows oxygen vacancies which were probably created during the BFCO deposition in a low oxygen pressure ambient (compared to the oxygen pressure during SRO deposition).

Figure 2A shows XRD θ-2θ scan of 150nm-thick BFCO films grown on epitaxial SRO films on (100)- and (111)-oriented STO substrates. The XRD spectra of all BFCO films show only 00*l* and *hhh* peaks and the corresponding peaks of the STO substrates, demonstrating that the BFCO films are well (001)- and (111)-oriented along the normal to the substrate surface. XRD analysis with extremely long counting times in the range 2θ= 91° - 97° were also performed in order to exclude the presence of magnetic parasite phases, as the ones reported by Bea et *al.*.[11] No peaks corresponding to α- or γ-$Fe_2O_3$ reflections were detected, even after these extra long scans. For these films grown on SRO/STO(100) and SRO/STO(111), the lattice parameter in pseudocubic symmetry calculated from the measured out-of-plane d-spacing was found to be 0.395 nm in both cases and no impurity phases were detected. Figure 2C displays Φ-scans of the BFCO 101 and STO 110 planes for a BFCO/SRO/STO(100) specimen. The four sharp peaks originate from BFCO, indicating that the film has good in-plane orientation. These four peaks are at the same positions as the four corresponding peaks from the STO substrate, demonstrating a cube-on-cube epitaxial growth of BFCO with respect to the substrate. This observation also indicates that the film has a four-fold symmetry axis (or a symmetry very close to it).

Accordingly, Figure 2D shows a Φ-scan of the BFCO 101 and STO 110 planes for (111)-oriented BFCO/SRO/STO(111) heterostructure. A three-fold symmetry is clearly seen, similar to the symmetry of the 111 surface of the substrate. The XRD pattern of (111)-oriented BFCO films (Fig. 2B) exhibit 111 and 333 reflection peaks of the double perovskite unit cell ($a_0$ ~ 0.79 nm), indicating a clear B-site $Fe^{3+}$ and $Cr^{3+}$ cation ordering, which has never been reported before for this material. If the indices of the pseudo-cubic *single* perovskite unit cell extending the STO lattice structure were used instead of those of the *double* perovskite unit cell, the 111 and 333 reflection peaks would be indexed $\frac{1}{2}\frac{1}{2}\frac{1}{2}$ and $\frac{3}{2}\frac{3}{2}\frac{3}{2}$, respectively, clearly corresponding to reflections from a superlattice having a periodicity equal to two lattice units in the corresponding direction. For the (001)-oriented film, the (00*l*) superlattice reflections of doubled perovskite unit cell with odd *l* numbers are not seen because of the extinction rule attached to the cubic face-centered symmetry. The presence of cation ordering demonstrates that the films have the predicted crystal structure.[16] It should be noted here that additional reflections (at 2Θ = 33.5° and 69.5°) corresponding to an unidentified contamination of the STO substrate (and therefore also present on the XRD spectra of bare substrates) are also visible. The pseudo-cubic lattice unit *a* = 0.395 nm calculated from the XRD data for both (001)- and (111)-oriented films is larger than that observed in polycristalline films, deposited on the same conditions on silicon substrates having 100 nm thick amorphous $SiO_2$ (pseudo-cubic lattice parameter ~ 0.394 nm). It is reflecting a compressive strain in the lateral direction owing to fully strained epitaxy of lattice mismatched BFCO on STO. This means that the lattice of the films having different orientations is deformed by the in-plane stress in a different manner and is therefore stabilized in different strained symmetries, tetragonal for the (001)-oriented films and rhombohedral for the (111)-oriented ones. Such differences in the crystal distortion therefore affect the strain-sensitive properties like ferroelectricity and magnetization in different manner.

Figure 3 shows a cross-sectional transmission electron microscopy image of an (001)-oriented epitaxial BFCO film observed along the [010] direction. The epitaxial growth of BFCO is clearly visible, with columnar epitaxial grains 100 nm - 150 nm in lateral size extending through the whole film thickness. Small angle grain boundaries are visible between the grains that are not disturbing the epitaxy. The surface of the film is flat and extremely smooth. The electron diffraction pattern confirms a very good epitaxy of the (001)-oriented BFCO layer, as already verified by $\Theta$-$2\Theta$ and $\Phi$-scans in XRD.

Super-lattice spots (¼ 0 ¼) are visible along the [1 0 -1] direction (best visible left from the (1 0 -1) index) (note indices used in the SAED patterns are the pseudo-cubic single perovskite BFCO half unit cell extending the STO substrate lattice). The presence of these super-lattice spots confirm the observations of super-lattice peaks made using X-ray diffraction. The super-lattice being observed along the [111] direction (XRD), being observed along the [1 0 -1] direction (SAED), and not being observed along the [100] direction (XRD & SAED) due to the extinction rules related to the cubic face centered symmetry, therefore confirms the assumed double perovskite crystal structure with a $Fe^{3+}$/$Cr^{3+}$ ordering along the [111] direction. The fact that (¼ 0 ¼) spots are observed, and not (½ 0 ½) superlattice spots like in the XRD spectra, could be a hint that regions with different ordering coexist, for instance regions with different $Fe^{3+}$/$Cr^{3+}$ ordering in a plane perpendicular to the [1 1 1] direction, that could change the local symmetry and give rise to such an effect. Further work is in progress to observe the $Fe^{3+}$/$Cr^{3+}$ ordering by electron diffraction (SAED) along other directions by preparing cross-sectional samples with other orientations.

Interestingly enough, although the Fe-Cr ordering is clearly demonstrated by the presence of a super-lattice revealed both by XRD and electron diffraction at the unit cell scale, the chemical composition is however subject to slight fluctuations at a longer scale as revealed by a chemical mapping of the cationic species present in the BFCO layer (Fig. 4). Such mapping shows that bismuth is uniformly distributed. The variations of the Bi signal are due to minor thickness variation of the cross-section TEM sample, as clearly shown when compared to the Sr and Ti signals from the single crystal STO substrate (note that the Sr layer in Fig. 4C is wider than the Ti layer in Fig. 4B, since it includes the Sr of the SRO electrode layer). There are, however, slight modulations in the Fe and Cr compositions as show in Fig. 4E and Fig. 4F. The Fe and Cr maps of the layer in the analyzed region exhibit a stripe pattern where the modulation width is in the order of tens of nm with Fe and Cr compositions being complementary to each other. As the chemical mapping at high spatial resolution is very sensitive to local modulations in composition, the stripe contrast shown in Figs. 4E & 4F suggests minor excess or depletion of Fe and Cr. This effect could be due to the presence of some longer range disorder within the Fe-Cr local ordering revealed above by diffraction techniques. The exact variation of Fe and Cr stoichiometry and the extent of the B-site cationic disorder are under further investigation in order to fully quantify these effects.

Let us now report on the physical properties of the epitaxial BFCO films. Piezoresponse force microscopy (PFM) was used to probe the ferroelectric properties of both the (001)- and (111)-oriented BFCO films. A positive voltage was applied between the conductive tip of the AFM and the bottom electrode (SRO) to pole the BFCO layer so that its spontaneous polarization points "down" and a square is written in the "down" state (black contrast in the right piezoresponse image of Fig. 5A). Next, a negative voltage is applied between the conductive tip of the AFM and the bottom electrode (SRO) to pole the BFCO layer so that its spontaneous polarization points "up" and a smaller square is written in the "up" state (white) within the first "down"-polarized (black) square.

Figure 5A shows the PFM images of the 150 nm (111)-oriented film before (left image) and after (right image) the "down" and "up" domains were written. We observe a clear contrast that reveals the presence of down and up ferroelectric domain in the BFCO thin film proving that these domains can be switched upon an application of an external voltage of the opposite polarity.

To further demonstrate the ferroelectric character of the BFCO films, i.e. a polarization that can be switched by the application of an electric field, local remanent piezoelectric hysteresis loops were measured by positioning the probing tip at a chosen site on the film surface and recording the piezoresponse signal as a function of an applied dc-voltage superimposed on the probing ac-voltage.[21] Figure 5B and 5C show piezoresponse hysteresis loops obtained for the (111)- and (001)-oriented BFCO films, respectively. The presence of these hysteresis loops confirms the presence of a switchable polarization, hence the ferroelectric character of BFCO thin films. Figure 5B and 5C show that piezoelectric properties are locally switched for both crystallographic orientations of the epitaxial BFCO films. Because the experiments were performed with different types of cantilevers and the cantilever response is dependent on its elastic properties, the magnitude of the signal in Fig. 5B and 5C cannot be compared. For the same reasons, an absolute value for the spontaneous polarization cannot be easily computed.[22,23] Nevertheless, from the squareness of the two loops we deduce that switching occurs differently for the two orientations: in (111) films, nucleation of opposite domains takes place at relatively low electric fields, but complete switching under the AFM tip requires a high field. In contrast, for (100)-oriented films domain nucleation occurs at moderate electric fields (around an effective value of the electric field of 330 kV/cm) but switching is then completed very fast.

Analyzing the size of the switched domains for each of the films (400 nm for the (111) film and 250 nm for the (100) oriented film), we infer that domains grow laterally more easily when a field is applied along the [111] direction than for a field along the [001] direction. The reversed polarization is stable, at least at the experiment time scale, several hours.

Let us now look at the magnetic properties of the BFCO epitaxial layers. Figure 6 shows magnetization hysteresis cycles M(H) at room temperature for two epitaxial 150 nm thick BFCO films having (001) and (111) crystallographic orientations. The magnetic field was applied in-plane, along [100]/[010] for the (001)-oriented films, and along [110] for (111)-oriented films. The component of the magnetic moment measured is parallel to the applied field. All three films appear to be reasonably saturated above 3kOe. It should be mentioned, however, that a very slow approach to saturation over an applied field range of several Tesla would be hard to detect with our system, which is limited to a maximum applied magnetic field of 2 Tesla. For 150 nm thick (111)-oriented BFCO, the saturation magnetization ($M_s$) was ~ 15 electromagnetic unit per cubic centimeter (1 (emu)/cc = 1kA/m in SI units) corresponding to a net magnetic moment (at 2T) of 0.2 $\mu_B$ per formula unit of BFCO.

On the other hand, the saturated magnetization ($M_s$) of a 150 nm (001)-oriented BFCO film was found to be ~30 emu/cc at room temperature, corresponding to ~0.4 $\mu_B$ per f.u. The results clearly reveal different intrinsic magnetic behavior for these BFCO thin films having different crystallographic orientations.

Indeed, the saturation magnetizations along these two crystallographic directions did not coincide at high saturating magnetic field, which would be the case if the difference in magnetization was solely due to crystal anisotropy. The fact that this was not observed for our epitaxial BFCO films of various crystallographic orientations indicates a fundamental difference in the *intrinsic* magnetic phase between these films. The stress and deformation, which are dependent on the thickness and crystallographic orientation of the epitaxial films, are expected to modify the magnetocrystalline and magnetoelastic anisotropy of the films and are likely to play an important role in the onset of the observed distinct magnetic phases.

**Conclusion**

In summary, using pulsed laser deposition, we have synthesized epitaxial thin films of $Bi_2FeCrO_6$ both on epitaxial $SrRuO_3$ on (100)- and (111)-oriented $SrTiO_3$ single crystalline substrates. In both cases, the $Bi_2FeCrO_6$ films have a double perovskite crystal structure very similar to that of BFO with a Fe-Cr ordering along the [111] direction. The films have been shown to have the correct cationic average stoichiometry throughout their thickness and no inclusions of magnetic iron oxide phases could be detected. A slight modulation in the Fe and Cr compositions forming stripe pattern where Fe and Cr compositions are complementary to each other, suggests minor excess or depletion of Fe and Cr on a 10 nm scale, possibly resulting from the presence of some longer range disorder within the Fe-Cr local ordering revealed by diffraction techniques.
The epitaxial BFCO films exhibit good ferroelectric and piezoelectric properties as well as magnetic properties *at room temperature*, a property that was not predicted by *ab-initio* calculations. The BFCO films also exhibit an unexpected crystallographic orientation dependence of their room temperature magnetic properties.
Our results qualitatively confirm the predictions made using the *ab-initio* calculations about the existence of good multiferroic properties in BFCO films, and reveal the unpredicted, yet promising, existence of magnetic ordering at room temperature. Studies investigating the magnetoelectric coupling in these BFCO films are currently underway.

**Acknowledgments**
The authors want to thank Prof. R. W. Paynter for insightful comments about X-ray photoelectron spectroscopy, Dr. C. Andrei for help in the analytical microscopy work, and Prof. James F. Britten for insightful discussions on X-ray diffraction. Part of this research was supported by NSERC (Canada), and FQRNT (Québec).

**FIGURE CAPTIONS**

**Figure1 A** XPS spectra at various depths from surface to interface of Fe2p and Cr2p core levels for a 300 nm thick BFCO film deposited on SRO on STO(100) crystalline substrate as a function of the depth from the surface. The same region of the spectra of a 300 nm thick BFO layer (Fe2p3/2 peak near 710.9 ± 0.2 eV) measured at a depth of 100 nm is also displayed for comparison. **B** EDXS in cross sectional TEM obtained from the same heterostructure.

**Figure 2** XRD Pattern for the 150nm-thick films and Φ-scan of the (101) plan of **A-C** grown on SRO on STO (100) and **B-D** on SRO on STO (111) substrates. Peaks due to the B-site ordering are seen in **B**. Because of extinction rule, these are not seen in **A**. Stars and squares represent the $K_\beta$ of STO and $W_{L\alpha}$ contamination, respectively.

**Figure 3** Cross-section TEM image showing a 115 nm thick epitaxial BFCO layer on an epitaxial SRO layer on STO (001). The direction of observation is [010]. The epitaxial growth is clearly visible, as are the smooth and flat BFCO surface and some low angle grain boundaries. **Inset:** The electron diffraction pattern demonstrates the good epitaxy of the BFCO layer and some (¼ 0 ¼) super-lattice spots are visible along the [1 0 -1] direction (best visible left from the (1 0 -1) index). The indices used are those of the pseudo-cubic single perovskite BFCO half unit cell extending the STO substrate lattice (see text for further details).

**Figure 4** XTEM overview and chemical mapping of Sr, Ti, Bi, Fe and Cr for a 115 nm thick BFCO epitaxial layer on an epitaxial SRO layer on STO (001). **A** Overview XTEM image in HA-ADF mode; the purple box delineates the area analyzed. **B** Ti chemical mapping using the Ti K$\alpha_1$ line. **C** Sr chemical mapping using the Sr K$\alpha_1$ line. **D** Bi chemical mapping using the Bi L$\alpha_1$ line. **E** Fe chemical mapping using the Fe K$\alpha_1$ line. **F** Cr chemical mapping using the Cr K$\alpha_1$ line. The chemical signal stronger in the middle of the image and weaker on the right side of the image is an artifact due to a thickness variation across the field of view.

**Figure 5 A** PFM images before and after writing square domains on the 150 nm thick BFCO (111) layer using the AFM tip as a top electrode. Piezoresponse signal as a function of dc bias voltage for **B** (111)- and **C** (001)-oriented BFCO films (~150 nm thick) grown on SRO (~30 nm) on STO substrates.

**Figure 6** Magnetic hysteresis loops measured by vibrating sample magnetometry (VSM). The two films appear to be saturated above 3 kOe.

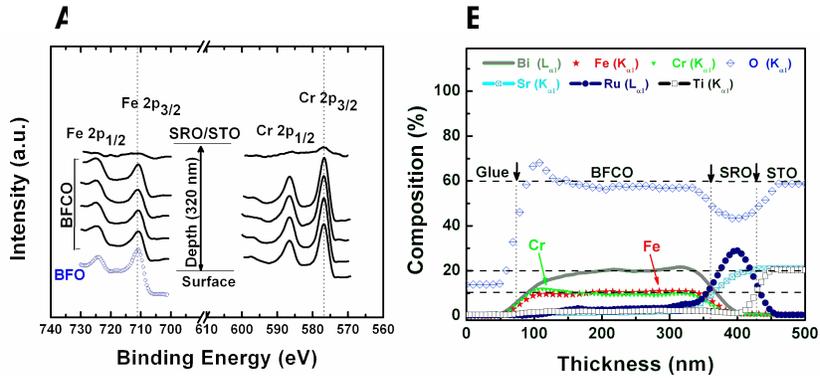

**Figure 1. R Nechache et al.**

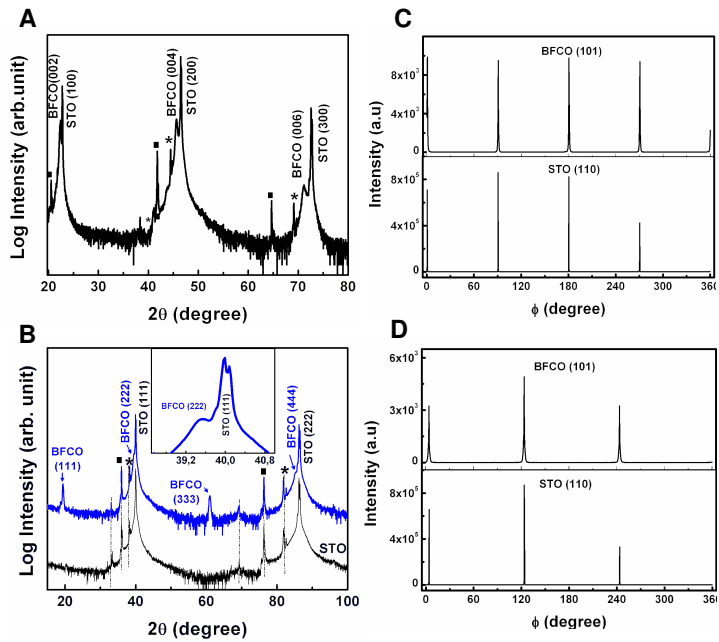

**Figure 2. R Nechache et al.**

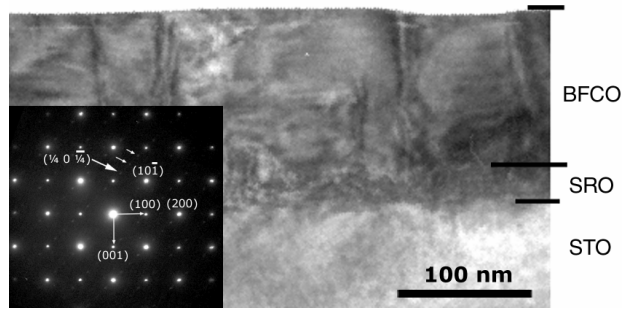

**Figure 3. R Nechache et al.**

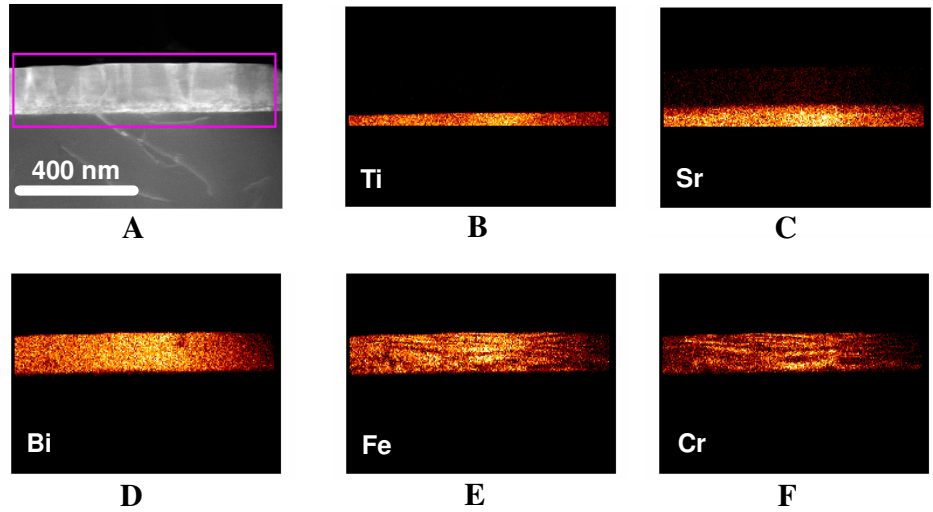

**Figure 4. R Nechache et al.**

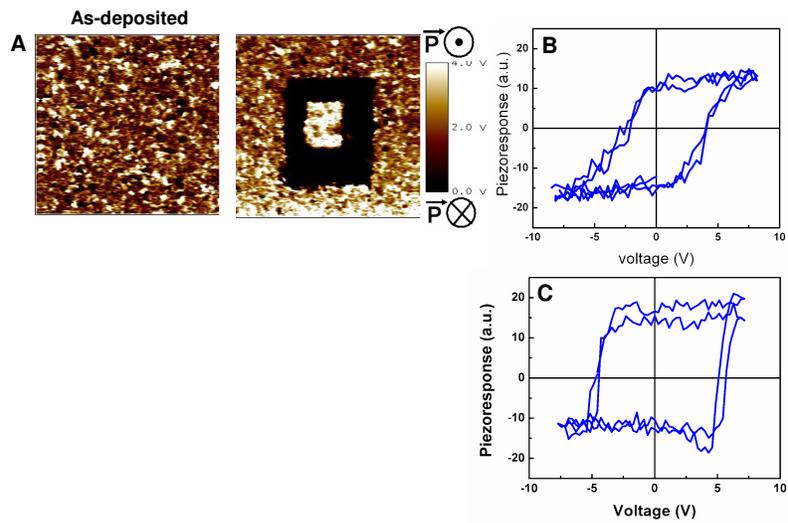

**Figure 5. R Nechache et al.**

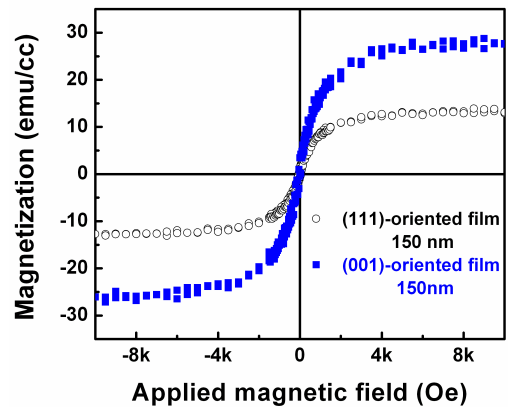

**Figure 6. R Nechache et al.**